# A fast approach to estimating Windkessel model parameters for patient-specific multi-scale CFD simulations of aortic flow

Zongze Li[1] and Wenbin Mao[1*]

[1]Department of Mechanical Engineering, University of South Florida, Tampa, Florida, USA

***Corresponding Author:**
Wenbin Mao

E-mail: wmao@usf.edu; Phone: 813-974-6636

Assistant Professor
Department of Mechanical Engineering
University of South Florida
4202 E. Fowler Avenue, ENG 030
Tampa, FL 33620

**Abstract:**

Hemodynamics in the aorta from computational fluid dynamics (CFD) simulations can provide a comprehensive analysis of relevant cardiovascular diseases. Coupling the three-element Windkessel model with the patient-specific CFD simulation to form a multi-scale model is a trending approach to capture more realistic flow fields. However, a set of parameters (e.g., $R_c$, $R_p$, and $C$) for the Windkessel model need to be tuned case by case to reflect patient-specific flow conditions. In this study, we propose a fast approach to estimate these parameters under both physiological and pathological conditions. The approach consists of the following steps: (1) finding geometric resistances for each branch using a steady CFD simulation; (2) using the pattern search algorithm to search the parameter spaces by solving the flow circuit system with the consideration of geometric resistances; (3) performing the multi-scale modeling of aortic flow with the optimized Windkessel model parameters. The method was validated through a series of numerical experiments to show its flexibility and robustness, including physiological and pathological flow distributions at each downstream branch from a healthy aortic geometry or a stenosed geometry. This study demonstrates a flexible and computationally efficient way to capture patient-specific hemodynamics in the aorta, facilitating the personalized biomechanical analysis of aortic flow.

1. Introduction

Cardiovascular diseases are the leading cause of death globally [1] and hemodynamics has a direct relation with the onset and development of cardiovascular diseases, such as hypertension, atherosclerosis, valvular heart disease, and heart failure [2-5]. However, comprehensive hemodynamics of the cardiovascular system can be difficult to measure due to the limitation of existing clinical tools. For instance, magnetic resonance imaging (MRI) and computed tomography (CT) has a lower temporal resolution [6-8]. Cardiac catheterization can evaluate cardiac function, but it is invasive and can only achieve the information within a limited region instead of comprehensive hemodynamics [9]. Doppler echocardiography has a high temporal resolution but cannot evaluate hemodynamics precisely. Hence, in recent decades, numerical approaches receive increasing attention due to the comprehensive and non-invasive features of obtaining hemodynamics. However, the astronomical amount of vasculature with intricate geometry plus the lack of computational performance prevent researchers from studying the hemodynamics of a whole vessel system but only allow them to simulate a specific section with interests. Therefore, proper boundary conditions at the openings of the geometry are critical to the accuracy of numerical simulations. Numerical modelling differs significantly across studies in terms of boundary conditions and may compromise the accuracy [10-13]. For example, constant (or zero) pressure outlet boundary conditions utilized in numerical simulations may induce large deviations in comparison to physiological data [13]. Therefore, lumped parameter models are coupled with downstream pressure outlet boundary conditions in computational fluid dynamics (CFD) to provide physiological hemodynamics, consequently leading the simulation results to be more reliable for clinical diagnosis [14].

The Windkessel model is one type of the lumped parameter model, which relates the pressure with the flowrate, and has been frequently used in CFD to couple with pressure outlet boundary conditions in aortic flows. The first Windkessel model was introduced by Otto Frank [15] in 1899 with two elements, a peripheral resistance ($R_p$) and a compliance ($C$) parallel to the resistance, in an electrical circuit analogy. However, the two-element model was not able to produce a realistic aortic pressure waveform due to the poor medium- to high-frequency representation of the systemic input impedance [16]. To deal with this shortage, the three-element Windkessel model was introduced [17], adding one characteristic impedance ($R_c$) series with the two-element Windkessel model. Although the three-element Windkessel model was able to provide a physiological pressure waveform from the flowrate waveform, the parameters deviated from the ones with physiological meaning. Therefore, the four-element Windkessel model was created to solve this problem, which has an inertia term ($L$) parallel to the characteristic impedance. However, the purpose of the Windkessel model coupled with CFD is only to provide time-dependent (varied by the flowrate) pressure boundary conditions for outlets to obtain physiological hemodynamics. Hence, the three-element Windkessel model (WK3) is mainly utilized in CFD and accordingly was applied in this paper.

To use the WK3, a set of three parameters ($R_c$, $R_p$, and $C$) for every outlet is necessary to be estimated for individual patients. Some algorithms have been investigated in this field to estimate these parameters. S. Pant et al [18] performed an unscented Kalman filter to find the optimized set and iterated periodic CFD simulations (i.e., unsteady blood flow over several cardiac cycles) to improve the set. Alimohammadi et al. [19] applied the data assimilation technique to search the set and improved the set with iterative periodic CFD simulations. Romarowski et al. [20] estimated the optimized set by least-square minimization. Some of the above algorithms involved periodic CFD simulation iteratively, which are time-consuming. They may also ignore the pressure drop between the inlet and each outlet when optimizing the WK3 parameters. There is one work done by Bonfanti et al. [21], of which the algorithm is able to control not only the extremum of pressure waveform of ascending aorta but also the flow distribution of downstream branch. However, the geometric resistances in their algorithm as well as the ones from S. Pant et al. [18] were approximated from a simplified analytical formula, which may not reflect the real resistance from a specific geometry and flow condition.

In this paper, we propose a fast algorithm to find the optimized set of parameters for the WK3 in aortic flows. Using this algorithm, the extremum of the pressure waveform at the ascending aorta and the flow distribution of each downstream branch can be controlled with little error. Furthermore, only one steady CFD simulation is involved in this algorithm, which significantly reduces the computational cost. The following content was organized as follows: The detailed procedure of the algorithm is introduced in the method section. Physiological and pathological flow conditions are analyzed to prove the flexibility and robustness of the algorithm in the result section. In the discussion section, the comparison with similar algorithms from other literature and additional results to support the algorithm are presented.

## 2. Method

### 2.1 CFD simulation setup

The blood was modeled as a Newtonian fluid with a density of 1060 kg/m$^3$ and a kinematic viscosity of 3.3e-6 m$^2$/s. The aortic geometry (shown in Fig. 1a) was adopted from an open-source library [22]. The aortic wall was defined as rigid with a no-slip boundary condition. The ascending aorta (AAo) utilized the velocity Dirichlet boundary condition with a uniform profile, imposing the flowrate waveform (shown in Fig. 1b) from "MICCAI CFD challenge 2012" [23] with a peak Reynolds number of 2799, a mean Reynolds number of 750, and a Womersley number of 23. The Reynolds number and Womersley number were calculated based on the diameter (2.82 cm) and the radius of AAo, respectively. The downstream branches were constrained by pressure boundary conditions with time-dependent pressure values calculated from the ordinary differential equation (ODE) of WK3 (shown in Eq. 8). A mesh independence study has been done (shown in Fig. 1c). When the lattice size is smaller than 0.02 cm, the mesh independence was achieved and the variation of the instant velocity magnitude at the center of RCCA at peak systole was within 2%. The overall number of lattices allocated in the flow domain was about 20.98 million.

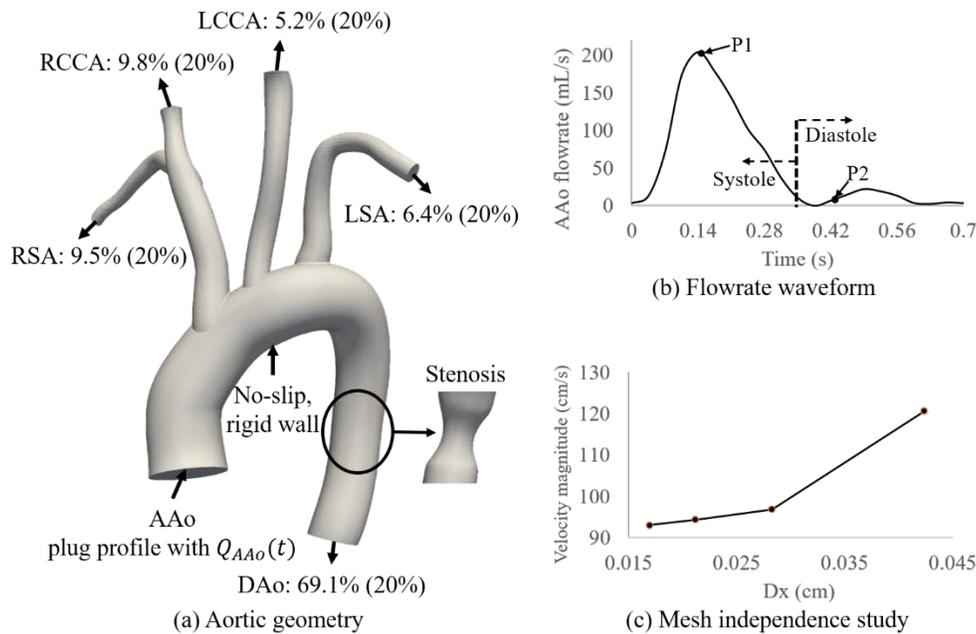

Figure 1. (a) Geometry used in the CFD simulation with the abbreviations of all six openings - AAo: Ascending Aorta; DAo: Descending Aorta; RCCA, LCCA: Right or Left Common Carotid Artery; RSA, LSA: Right or Left Subclavian Artery. The percentage after the colon denotes the physiological and even (20% for each downstream branch) flow distributions used in the result section. Black circle indicates the location of stenosis artificially modified from the original geometry and the section presented right to the geometry shows the shape of stenosis. Except for the stenosed location, the other region is identical to the original geometry. (b) The waveform of flowrate $Q_{AAo}(t)$ used for the velocity boundary condition with the plug flow profile at the AAo. P1 (0.15 s) and P2 (0.44 s) are the peak systole and the early diastole time

points, respectively. The vertical dot line located at 0.35 s separates systole and diastole. (c) The result of the mesh independence study was based on a periodic CFD simulation with zero pressure outlets and a Poiseuille profile velocity inlet with the value from $Q_{AAo}(t)$. The velocity magnitude of the center point on the cross-section of RCCA at peak systole was used as the check value.

**2.2 Lattice Boltzmann method**

The lattice Boltzmann method (LBM) has been an alternative and promising numerical method for simulating fluid flows and modeling physics in fluids, especially succeeding in the case involving interfacial dynamics and complex geometry [24]. The fundamental idea of LBM is to have simplified kinetic models that incorporate microscopic processes with macroscopic values since the macroscopic values are the collective behavior of many microscopic particles in the system [25]. In the LBM, the fluid domain is voxelized into structured lattices. Each of them has its own distribution function $f(\vec{x}, \vec{c_i}, t)$, describing the mass density of fluid particles with velocity $\vec{c_i}$ at lattice node $x$ and time $t$. In each lattice, the particle velocity in the distribution function is also discretized. The most widespread lattice structure in 3D simulations is D3Q19 stencil, which means each lattice node has 18 neighbor nodes with a total of 19 lattice velocities. In this study, a D3Q19 lattice was used. Through the Chapman-Enskog expansion, in a nearly incompressible flow limit, the Navier-Stokes equations (N-S) governing fluid flow can be restored with second-order accuracy in space and time [26]. The solving process of the discretized Boltzmann equation involves two steps: the collision step and the streaming step (the propagation of distribution function to its neighboring lattices). The time evolution of the particle distribution functions takes the form of Eq. 1.

$$f_i(\vec{x} + \vec{c_i}\Delta t, t + \Delta t) = f_i(\vec{x}, t) + \Omega_i(\vec{x}, t) \tag{1}$$

In the equation, $i$ denotes the lattice direction ($i = 0, 1, \ldots, 18$ for D3Q19), $\Delta t$ is the time increment, $\vec{x} + \vec{c_i}\Delta t$ gives the neighbor lattice location on the $\vec{c_i}$ direction, and $\Omega_i(\vec{x}, t)$ is the collision term. In this study, the Bhatnagar-Gross-Krook (BGK) collision operator is applied, which takes the form of Eq. 2.

$$\Omega_i(\vec{x}, t) = -\frac{f_i(\vec{x},t) - f_i^{eq}(\vec{x},t)}{\tau} \Delta t \tag{2}$$

where $\tau$ is the relaxation time, which can be related to kinematic viscosity by Eq. 3, $c_s$ represents the speed of sound, which also determines the relation $p = c_s^2 \rho$ between pressure p and density $\rho$. In the isothermal collision operator (for example BGK), it is equal to $\sqrt{\frac{1}{3}\frac{\Delta x}{\Delta t}}$.

$$\tau = \frac{\nu}{c_s^2} + \frac{\Delta t}{2} \tag{3}$$

The equilibrium distribution function $f_i^{eq}(\vec{x}, t)$ is given by Eq. 4.

$$f_i^{eq}(\vec{x}, t) = w_i \rho [1 + \frac{\vec{u} \cdot \vec{c_i}}{c_s^2} + \frac{(\vec{u} \cdot \vec{c_i})^2}{2c_s^4} + \frac{\vec{u} \cdot \vec{u}}{2c_s^2}] \tag{4}$$

where $w_i$ is the weighting function of each lattice direction, $\rho$ and $\vec{u}$ are the macroscopic density and velocity of the current lattice that can be obtained from Eq. 5.

$$\begin{aligned} \rho(\vec{x}, t) &= \sum_i f_i(\vec{x}, t) \\ \vec{u}(\vec{x}, t) &= \frac{\sum_i \vec{c_i} f_i(\vec{x}, t)}{\rho(\vec{x}, t)} \end{aligned} \tag{5}$$

The viscous stress tensor $\boldsymbol{\sigma}$ can be calculated from the distribution function by Eq. 6, where $\alpha$ and $\beta$ are indices of x, y, and z directions, $f_i^{neq} = f_i - f_i^{eq}$.

$$\sigma_{\alpha\beta} \approx (\frac{\Delta t}{2\tau} - 1) \sum_i c_{i\alpha} c_{i\beta} f_i^{neq} \tag{6}$$

The wall shear stress (WSS) vector can be calculated using the viscous stress tensor as Eq. 7, where $\vec{n}$ is the unit normal vector of an arbitrary surface.

$$WSS = \boldsymbol{\sigma} \cdot \vec{n} - [(\boldsymbol{\sigma} \cdot \vec{n}) \cdot \vec{n}]\vec{n} \tag{7}$$

In this study, we build our modules on top of an open-source LBM library - Palabos [27]. Palabos is a software library developed since 2010 and has been used widely in different communities, including RBC flows [28], porous media [29], aeroacoustics [30], to name a few. The library was written in C++ and based on MPI for parallel executions.

**2.3 Three-element Windkessel model (WK3)**

The governing equation of WK3 (module 2 of Fig. 2a) takes the form of Eq. 8.

$$\frac{dP(t)}{dt} = \frac{R_c+R_p}{R_p C} Q(t) + R_c \frac{dQ(t)}{dt} - \frac{P(t)-P_{ref}}{R_p C} \tag{8}$$

where $R_c$ denotes the characteristic impedance, $R_p$ represents the flow resistance from downstream vascular system, $C$ is the total compliance due to the elasticity of downstream vascular system, and $P_{ref}$ is the reference pressure.

In the CFD simulations, the values of pressure boundary conditions for each downstream branch were updated every time step according to Eq. 8. In the equation, $R_c$, $R_p$, and $C$ were estimated from the proposed algorithm in section 2.4. The ODE was solved by a 4$^{th}$ order Runge-Kutta method, where the flowrate was measured at the current time step from each downstream branch and the time derivative of flowrate was calculated by 1$^{st}$ order backward Euler scheme from the measured flowrate.

**2.4 Estimation of WK3 parameters**

The goal of our proposed approach is to control the extremum of the pressure waveform at the AAo (i.e., $P_{max}$, $P_{min}$) and the flow distribution ($Q\%_{RSA}$, $Q\%_{RCCA}$, $Q\%_{LSA}$, $Q\%_{LCCA}$, $Q\%_{DAo}$) at each outlet. The flow distribution is defined as the ratio of flow volume going out at each downstream branch to the inlet flow volume in a cardiac cycle. To reach these target values, WK3 parameters must be tuned to obtain the desired results. Hence, in the following procedure, we assume the target values are known. The maximum and minimum pressures at the AAo were set to be 120 and 80 mmHg, respectively. Two sets of flow distributions were considered in this study (shown in Fig. 1a) and the flowrate waveform at the AAo is displayed in Fig. 1b.

Originally, the WK3 was considered at the root of the aorta (the location of the AAo in this paper) [31] in clinical studies to provide peripheral load to interpret the functionality of the experimental object [32] or related them to pathological states of human's body [33, 34]. However, in the CFD field, people couple this model with outlet pressure boundary conditions to achieve physiological pressure and velocity fields. In other words, this model is no longer applied at its original location. Hence, we introduce a geometric resistance $R_{g_i}$ for each downstream branch to represent the flow resistance of the 3D aortic geometry, where the subscript i denotes each downstream branch. The $R_{g_i}$ can be calculated by Eq. 9, where $P_{in}(t)$ and $P_{out_i}(t)$ are the pressure waveform of the AAo and each outlet branch, respectively. $Q_{out_i}(t)$ is the flowrate waveform of each downstream branch, and $T$ is the duration of a cardiac cycle. In steady CFD simulations, time average values in Eq. 9 can be replaced by steady-state values. With the consideration of $R_{g_i}$, the analogy of the geometry plus the WK3 is shown in Fig. 2a with module 2 (replace $M_i$ to module 2 in each downstream branch). Fig. 2b depicts the whole procedure to estimate the WK3 parameters. Firstly, a set of $R_{g_i}$ is calculated from a steady CFD simulation to represent the flow resistance of each branch. Secondly, they are considered in a circuit analogy to estimate the WK3 parameters by a global optimization algorithm. Finally, using the optimized WK3 parameters, we conduct periodic CFD simulations coupled with the WK3.

$$R_{g_i} = \frac{\int_0^T [P_{in}(t) - P_{out_i}(t)]dt}{\int_0^T Q_{out_i}(t)dt} \tag{9}$$

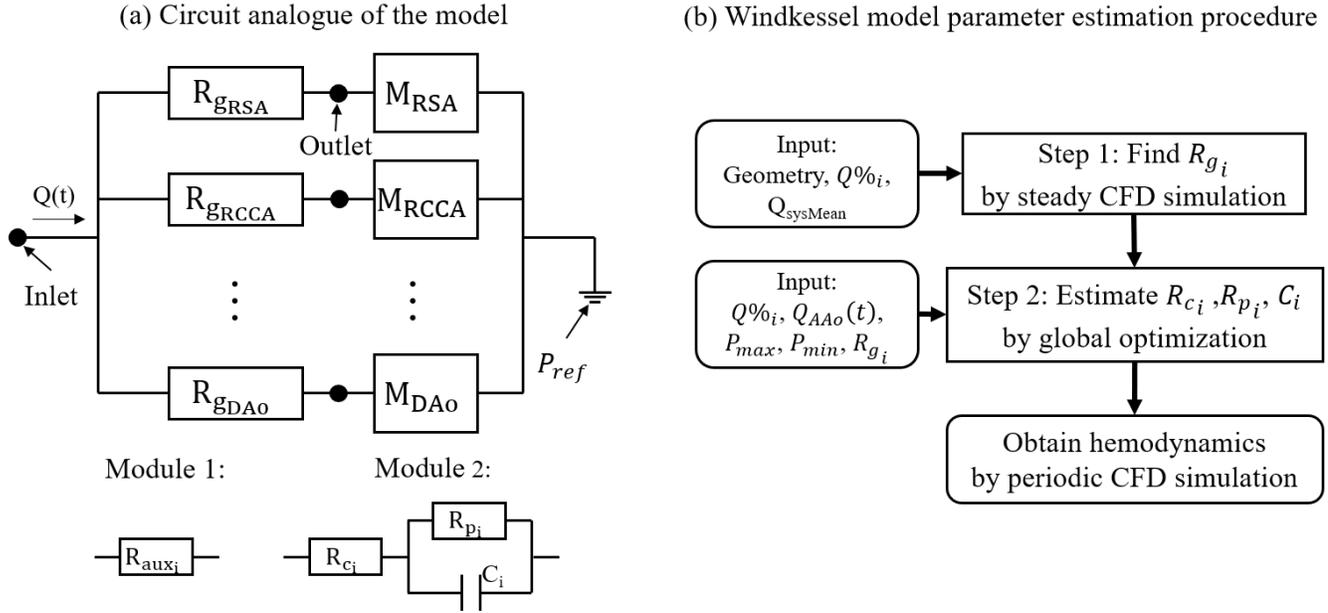

Figure 2. (a) Circuit analogy of the model: when this circuit is used in finding geometric resistance, $M_i$ is replaced by module 1. When it is used in estimating the WK3 parameters, $M_i$ is replaced by module 2. (b) The flowchart of the procedure to conduct a physiological CFD simulation. $Q_{sysMean}$ denotes the mean flowrate during the systole of a cardiac cycle.

The first step is to find $R_{g_i}$ of each downstream branch. There are two factors determining the $R_{g_i}$: the flow distribution and the geometry of the aorta. For a fixed geometry, to achieve target flow distribution, an auxiliary resistance ($R_{aux_i}$) is introduced after the $R_{g_i}$ to control the overall resistance of each branch, as shown in Fig. 2a by replacing $M_i$ with module 1. In this circuit analogy, if $R_{aux_i} \gg R_{g_i}$, the flow distribution relationship will follow Eq. 10. Note that only the ratio of $R_{aux_i}$ is fixed in this equation, one can choose any set of $R_{aux_i}$ which satisfies that $R_{aux_i}$ is large enough to dominate the overall resistances of each downstream branch ($R_{g_i} + R_{aux_i}$). Then, a steady CFD simulation is conducted with a proper set of $R_{aux_i}$ coupled with all the outlets and the mean flowrate during systole as the inlet boundary condition. If a proper set of $R_{g_i}$ is chosen in the simulation, the flow distribution should reach the target one and the pressure waveforms ($P_{in}$, $P_{out_i}$) and the flowrate waveforms ($Q_{out_i}$) from this steady CFD simulation can be used to calculate $R_{g_i}$ by Eq. 9. More detail about the reason to use this choice (the steady CFD simulation with the flowrate during systole) can be found in the discussion section.

$$Q\%_{RSA} : Q\%_{RCCA} : \dots : Q\%_i \approx \frac{1}{R_{aux_{RSA}}} : \frac{1}{R_{aux_{RCCA}}} : \dots : \frac{1}{R_{aux_i}} \tag{10}$$

The second step is to estimate the WK3 parameters with the consideration of $R_{g_i}$. In Fig. 2a with module 2, the flow distribution is also governed by a similar equation as Eq. 10, but the $R_{aux_i}$ become total resistance $R_{tot_i}$ which is defined as the sum of $R_{g_i}$ from the geometry and $R_{WK3_i} = R_{c_i} + R_{p_i}$ from the WK3, as shown in Eq. 11.

$$Q\%_{RSA} : Q\%_{RCCA} : \dots : Q\%_i \approx \frac{1}{R_{tot_{RSA}}} : \frac{1}{R_{tot_{RCCA}}} : \dots : \frac{1}{R_{tot_i}} \tag{11}$$

where $R_{tot_i}$ can be calculated by Eq. 12, $T$ is the duration of a cardiac cycle and the formula to calculate $P_{mean}$ is an empirical equation from Alimohammadi et al [19].

$$R_{tot_i} = \frac{P_{mean} - P_{ref}}{Q_{mean_i}}$$

$$P_{mean} = P_{min} + \frac{1}{3}(P_{max} - P_{min}) \tag{12}$$

$$Q_{mean_i} = \frac{\int_0^T Q(t)dt}{T} Q\%_i$$

Then the resistance and pressure for the WK3 of each downstream branch can be obtained from Eq. 13 and 14. Note that, $P_{max}$, $P_{min}$, and $P_{mean}$ in Eq. 12 are the extremum pressures at the AAo (inlet point in Fig. 2a) and the $R_{WK3_i}$, $P_{max_i}$, and $P_{min_i}$ are the values used in the search of WK3 parameters.

$$R_{WK3_i} = R_{tot_i} - R_{g_i} \tag{13}$$

$$P_{max_i} = P_{max} - R_{g_i} Q_{mean_i}$$
$$P_{min_i} = P_{min} - R_{g_i} Q_{mean_i} \tag{14}$$

After the $R_{WK3_i}$, $P_{max_i}$, and $P_{min_i}$ for the WK3 of each downstream branch are found, the pattern search algorithm [35] from the global optimization toolbox in MATLAB [36] is applied to find the optimized WK3 parameters for each downstream branch. The pattern search finds the optimized parameter set of an ODE to minimize the objective function. For example, in this study, the ODE is Eq. 8, where $Q_i(t)$ and $\frac{dQ_i(t)}{dt}$ are known values, $Q_i(t)$ is calculated by Eq. 15 for each downstream branch and $\frac{dQ_i(t)}{dt}$ can be obtained from $Q_i(t)$. We want to find an optimized set of $R_{c_i}$, $R_{p_i}$, and $C_i$ that can derive a pressure waveform of which the extremum values are equal to $P_{max_i}$ and $P_{min_i}$. Hence, $R_{c_i}$ and $C_i$ are the search parameters for the pattern search algorithm with the range of 0 to $R_{WK3_i}$ and 0 to 3 mL/mmHg respectively. The range of $C_i$ covers the compliance values used in the literatures [16, 19, 34, 37-39]. $R_{p_i}$ can be obtained by subtracting $R_{c_i}$ from $R_{WK3_i}$. In each search, the ODE is solved for 20 cardiac cycles to eliminate the transient effect from initial value and ensure to reach the periodic pressure waveform. The extremum pressure values ($P_{max_{ODE}}$ and $P_{min_{ODE}}$) in the last period of the resultant pressure waveform are recorded to calculate the objective function defined by Eq. 16. In this procedure, five sets of WK3 parameters for each downstream branch can be found and then used in the periodic CFD simulation to provide patient-specific boundary conditions. The procedure of step 2 is presented in Fig. 3.

$$Q_i(t) = Q_{AAo}(t) * Q\%_i \tag{15}$$

$$obj = \sqrt{\frac{1}{2}[(P_{max_i} - P_{max_{ODE}})^2 + (P_{min_i} - P_{min_{ODE}})^2]} \tag{16}$$

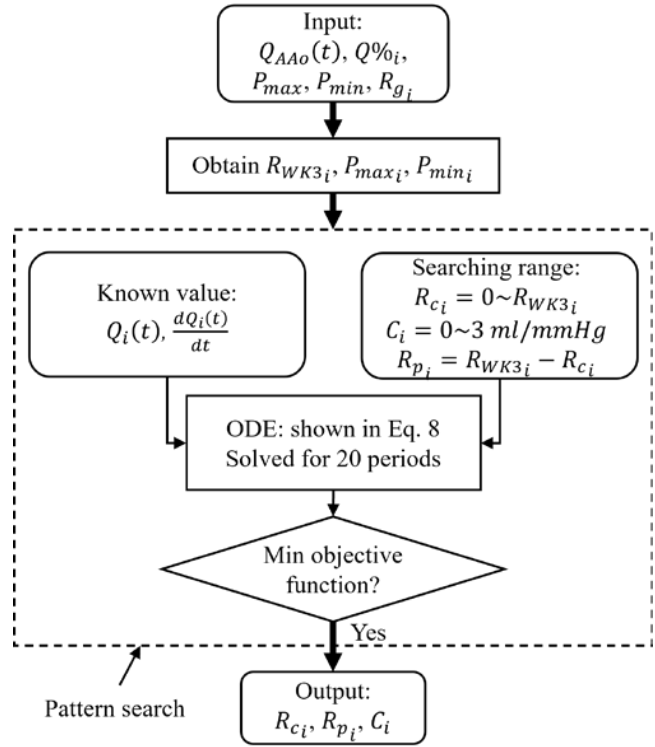

## 3. Result

### 3.1 Algorithm Robustness

To investigate the robustness of the proposed algorithm, three simulations with different geometries and flow distributions are studied and discussed. Fig. 1a demonstrates the geometry and flow distribution applied for three cases: (1) the physiological aortic geometry and flow distribution [40]; (2) the physiological flow distribution and a modified aortic geometry with stenosis at the DAo branch (the cross-section area is about one-third of the original one); (3) the physiological geometry with an evenly distributed flow at each outlet. The geometry with stenosis was modified by the open source software "Blender" [41]. All other parameters (e.g., inlet flow waveform and extremum of inlet pressure waveform) were fixed. The following content will present the results based on these three cases.

The $R_{g_i}$ and WK3 parameters of all three cases are summarized in Table 1. The $R_{g_{DAo}}$ in Case 2 increases about four times of the one in Case 1 due to the stenosis. The $R_{g_i}$ of the other branches varies slightly. However, the variation of the $R_g$ is negligible in comparison to the $R_c + R_p$ of the WK3, consequently leading to little difference between Cases 1 and 2 in WK3 parameters. As for Case 3 of an evenly distributed flow, except for the DAo (with a decreased flowrate), the $R_{g_i}$ of other branches raise due to the increased flowrate and the resistances of the WK3 reduce on account of the variation of the flow distribution. Note that, both the CFD simulation and the estimation of Windkessel model parameters were conducted with a reference pressure of 70 mmHg, which makes the relative pressure waveform oscillating from 10 to 50 mmHg instead of 80 to 120 mmHg. Hence, the optimized WK3 parameters may be different from other literature [16, 19, 34, 37-39]. This approach can decrease the time needed to achieve the convergence in the periodic CFD simulation coupled with the WK3. In this study, all the periodic CFD simulations use four cycles to reach the convergence and the last cycle is used to analyze the results in the following content. More detail about the comparison between reference and absolute pressure can be found in the discussion section.

Table 1. Geometric resistance and WK3 parameters obtained from the proposed algorithm. The three values from left to right in each table element are the parameters of Case 1, 2, and 3, respectively.

| Branch | $R_g$(mmHg*s/mL) | $R_c$(mmHg*s/mL) | $R_p$(mmHg*s/mL) | $C$(mmHg/mL) |
|---|---|---|---|---|
| DAo | 0.007\|0.029\|0.008 | 0.263\|0.262\|0.909 | 0.347\|0.325\|1.215 | 1.429\|1.412\|0.415 |
| RSA | 0.218\|0.217\|0.345 | 1.907\|1.907\|0.892 | 2.363\|2.363\|0.894 | 0.194\|0.194\|0.381 |
| RCCA | 0.047\|0.050\|0.077 | 1.855\|1.855\|0.907 | 2.449\|2.445\|1.147 | 0.203\|0.203\|0.411 |
| LCCA | 0.040\|0.044\|0.070 | 3.497\|3.497\|0.907 | 4.662\|4.657\|1.155 | 0.108\|0.108\|0.411 |
| LSA | 0.098\|0.100\|0.176 | 2.839\|2.839\|0.903 | 3.724\|3.722\|1.053 | 0.132\|0.132\|0.402 |

Table 2. Flow distribution, extremum of pressure waveform, and $L^2$ norm of error from the simulation results of all the three Cases.

| P unit: mmHg | Case 1 | Case 2 | Case 3 |
|---|---|---|---|
| DAo Q% | 68.81% | 68.61% | 20.38% |
| RSA Q% | 9.36% | 9.42% | 19.98% |
| RCCA Q% | 9.69% | 9.75% | 19.97% |
| LCCA Q% | 5.17% | 5.21% | 20.13% |
| LSA Q% | 6.33% | 6.37% | 19.79% |
| $L^2$ of Q% | 0.16% | 0.22% | 0.20% |
| AAo $P_{max}$ | 119.56 | 122.75 | 125.64 |
| AAo $P_{min}$ | 79.60 | 78.76 | 78.49 |
| $L^2$ of P | 0.42 | 2.14 | 4.13 |

Table 2 shows the $L^2$ norm errors of flow distribution and pressure extremum of three cases. The error is calculated by Eq. 17, where i denotes the downstream branch. $y_{i_{true}}$ is the target value (i.e., flow distribution percentage or extremum of pressure) and $y_{i_{pred}}$ is obtained from the simulation results.

$$L^2 \ norm \ error = \sqrt{\frac{1}{N}\sum_{i=1}^{N}(y_{i_{true}} - y_{i_{pred}})^2} \tag{17}$$

We can see this algorithm can handle situations with different geometries or abnormal flow distribution with relatively small error. The flow distribution can be controlled with high accuracy. As for the pressure, Case 1 and Case 2 provide good predictions with the errors of 0.42 and 2.14 mmHg respectively, while the error is slightly larger for Case 3.

### 3.2 Impact of stenosis and flow distribution on flow field

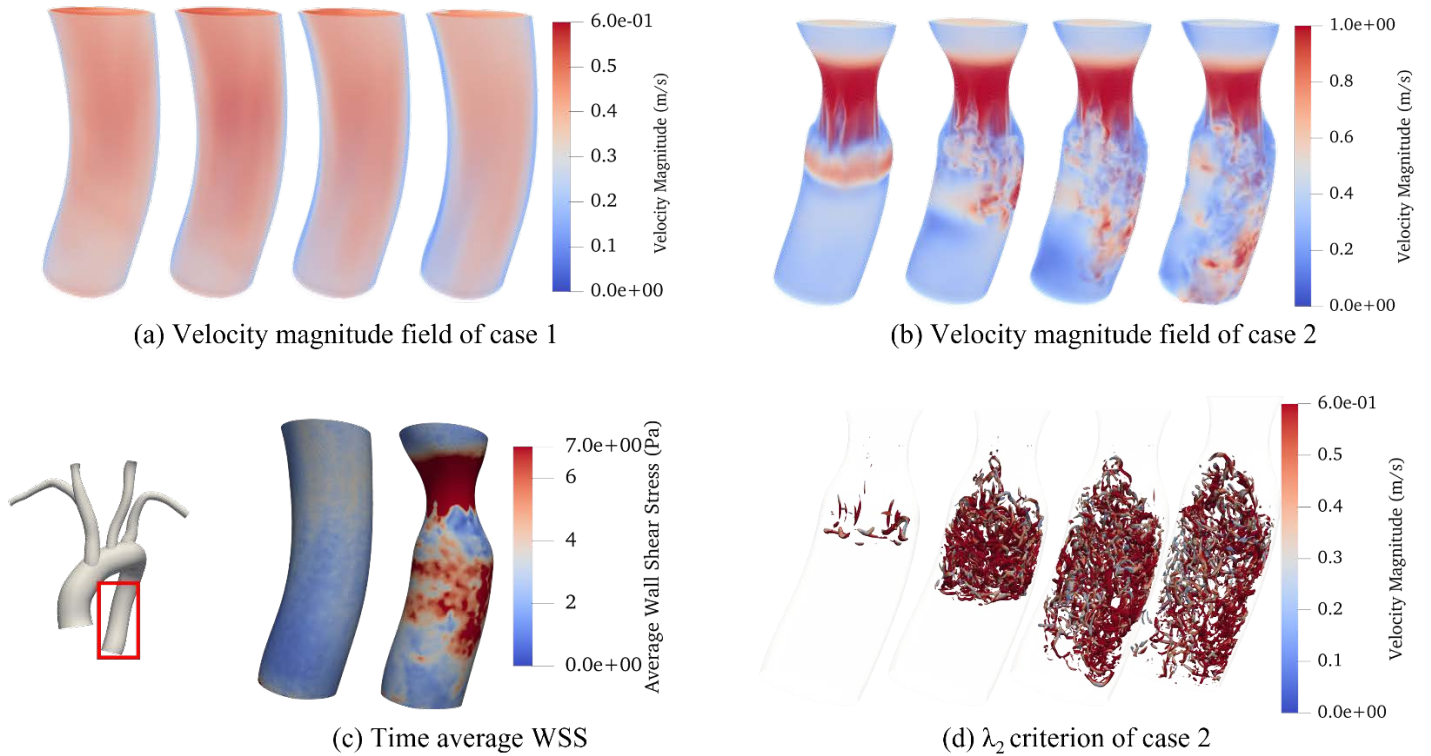

(a) Velocity magnitude field of case 1

(b) Velocity magnitude field of case 2

(c) Time average WSS

(d) $\lambda_2$ criterion of case 2

Figure 4. Results of four consecutive time points (0.131, 0.153, 0.175, and 0.197s) near peak systole are presented in figure a, b, and d. The left bottom inset shows the cutting region of the geometry presented in other figures. (a) Velocity magnitude field of Case 1 at the DAo branch. (b) Velocity magnitude field of Case 2. (c) Time average WSS (averaged from four consecutive time points mentioned above). (d) Vortices generated from the stenosed region in Case 2 (visualized by an iso-surface of $\lambda_2$ criterion of -400,000 s$^{-2}$).

Results of Case 1 and Case 2 are compared near peak systole (0.131 to 0.197 s) to reveal the impact of stenosis on the flow field. Fig. 4 demonstrates the velocity magnitude fields, time average WSS and $\lambda_2$ criterion. The value of iso-surface of $\lambda_2$ criterion is -400,000 s$^{-2}$. Case 1 does not have any vortex under the iso-surface of this value, so only the $\lambda_2$ criterion of Case 2 is presented. As shown in Fig. 4a, the aortic flow in the normal geometry does not generate any vortices in the branch of the DAo near peak systole. In contrast, the stenosis in Case 2 induces a jet at peak systole, agitates the downstream fluid, and generates complex vortices after peak systole (Fig. 4b and d). The maximum velocity magnitude is increased from 0.57 to 1.60 m/s (a 181% increase) by the stenosis. As shown in Fig. 4c, the stenosis also lead to the elevated temporal and spatial average wall shear stress from 2.17 to 6.10 Pa (a 281% increase), which could be related to the generation and rupture of the aneurysm [42]. Although, the stenosis in the branch of the DAo generates a central jet with increased velocity and WSS at peak systole, the upstream region of stenosis does not differ significantly from Case 1 as shown in Fig. 5 and 6. In Fig. 5, it is obvious that only the proximal region of some downstream branches has minor difference, for example, the LSA and RSA. At diastole (Fig. 6), except for the main aortic arch, the other downstream branches also present similar flow patterns. Hence, in general, the existence of stenosis does not significantly affect the upstream region if the flow distribution is the same.

Cases 1 and 3 are compared to analyze how different flow distribution affects the flow field. Since a higher flow distribution is specified in Case 3 on each downstream branch except for the DAo, higher velocity exists in these branches and an individual colormap is used to better present the flow pattern in Fig. 5 and 7. At peak systole, for Case 3, the velocity magnitude in the DAo region is low and a recirculation flow pattern exists downstream the bend of the LSA branch. Although the change in flow distribution increases the mean velocity of downstream branches (except for DAo), the flow pattern is similar at the peak systole. In contrast, at early diastole (shown in Fig. 6), the flow pattern in each branch is largely altered. The deceleration of flow generated many small vortices in the branches, reflecting a disturbed turbulent flow field. Vortices prevailed in the distal DAo branch in Case 1 due to stronger systolic flow, while no vortices are observed in the same region of Case 3. In general, flow distribution affects the velocity magnitude but has a minor influence on the flow pattern during systole. However, the difference in flow pattern during diastole is obvious. In *in vivo* cases, the flow distribution differs from patients and abnormal flow distributions might be related to diseases such as stenosis [43]. Hence, to achieve reliable velocity and pressure fields, considering the subject-specific flow distribution in CFD simulations of aortic flow is important.

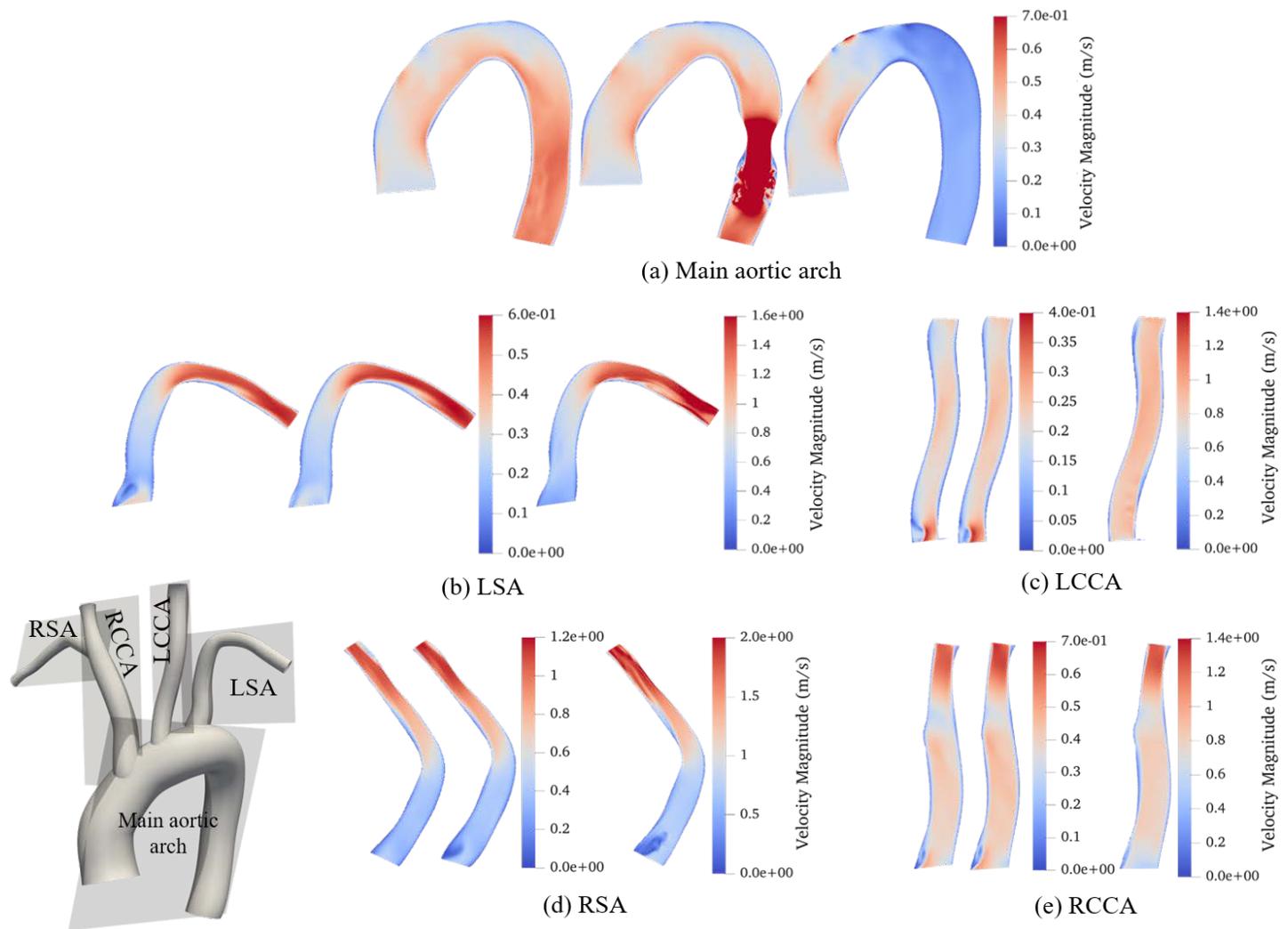

Figure 5. Velocity magnitude contours of each branch at peak systole (t = 0.15s) of all the cases, from left to right shows Case 1, Case 2, and Case 3, respectively. The left bottom inset shows the cutting planes of each branch for Fig. 5 and 7.

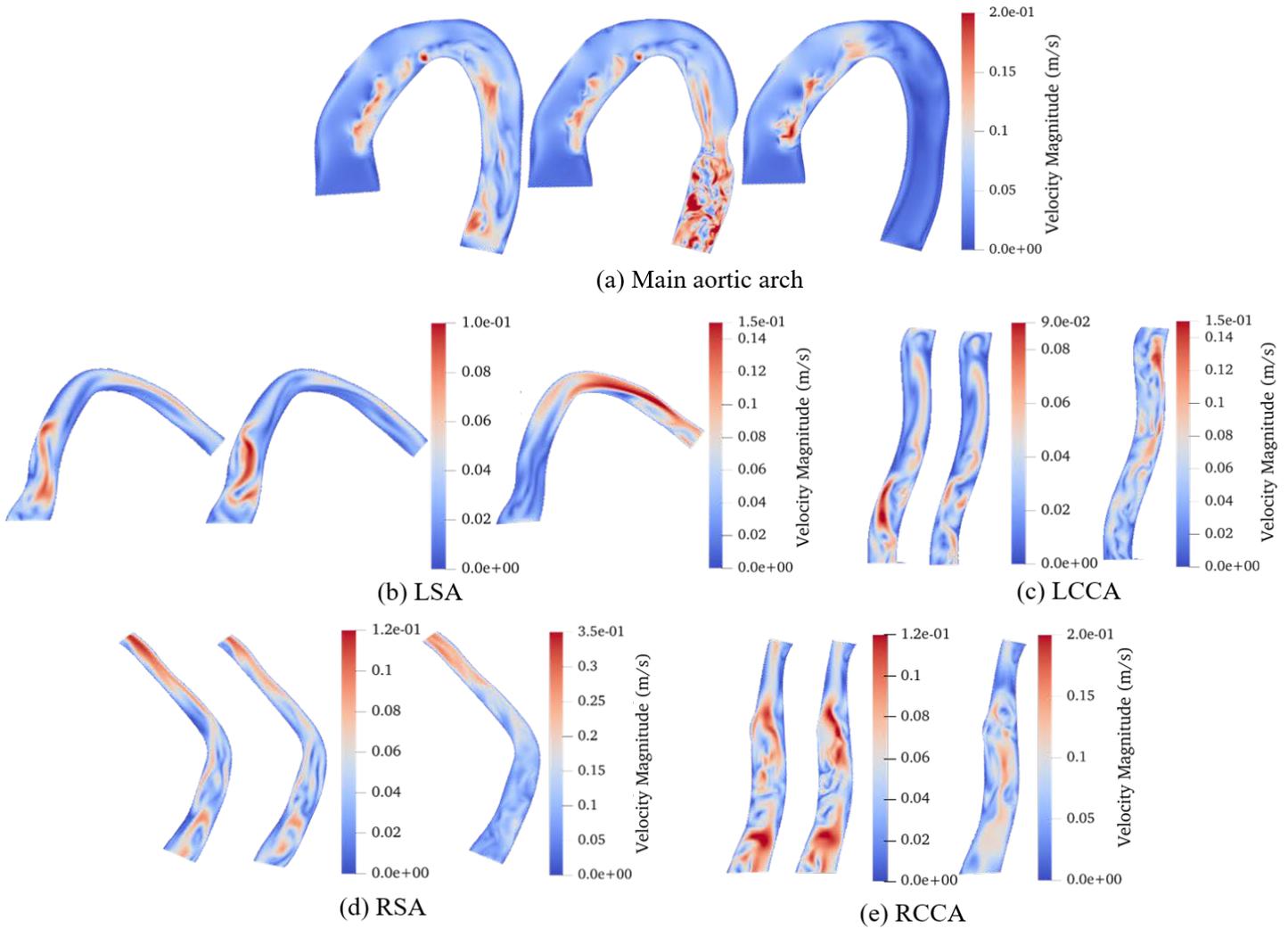

Figure 6. Velocity magnitude contours of each branch at early diastole (t = 0.44s) of all the cases, from left to right shows Case 1, Case 2, and Case 3, respectively.

## 4. Discussion:

In this paper, a fast approach to finding optimized parameters of WK3 coupled with CFD simulations of aortic flow has been proposed. Firstly, a set of geometric resistances which will be used in the estimation of WK3 parameters were obtained by a steady CFD simulation with a mean flowrate during systole as the inlet boundary condition. Secondly, the optimization of WK3 parameters for each downstream branch with the consideration of geometric resistances was achieved using a global optimization algorithm. Lastly, after getting the WK3 parameters of each downstream branch, a periodic CFD simulation was performed to obtain the patient-specific hemodynamics. The presented algorithm allows the control of the flow distribution of each outlet and the extremum of the inlet pressure waveform with a known flowrate waveform of the inlet. With the controllable flow distribution and extremum of pressure waveform, the outlet boundary conditions can be constrained according to subject-specific flow conditions. Consequently, more realistic velocity and pressure fields as well as the WSS can be obtained from CFD simulations.

In recent decades, several researchers have studied the optimization of WK3 parameters when coupling with CFD simulations for cardiovascular flows. Romarowski et al. [20] utilized a least-square method to find the optimized set of WK3 parameters. In their algorithm, no CFD simulation was conducted in the estimation of WK3 parameters. A least-square method was applied to achieve, by enumerating WK3 parameters, the minimum mismatch between the pressure waveform obtained from the ODE of WK3 and the target one, which was either from a clinical measurement or calculated from a set of empirical equations. Alimohammadi et al. [19] applied a method involving periodic CFD simulations to find the optimized set of parameters. Instead of considering geometric resistance, they directly used the optimized set obtained from the ODE iteration to run a periodic CFD simulation. The flowrate of each downstream branch of the periodic CFD simulation was then input back into the ODE iteration again to find the optimized WK3 parameters which were used for another periodic CFD simulation. This iterative procedure stopped until the objective function did not reduce any more. This method involved the iteration of periodic CFD simulations, which is time-consuming. Meanwhile, they both ignored the pressure drop between the inlet and each outlet in the circuit analogy when solving the ODE of WK3. Although under many situations, the pressure drop between the inlet and each outlet is quite small, they may also vary due to the difference in geometry, causing each outlet has a different pressure waveform which is used for the estimation of WK3 parameters. For example, if a severe stenosis exists upstream, it could largely increase the geometric resistance, inducing a large pressure drop in the branch. It could also change the flow distribution, consequently influencing the entire hemodynamics. Hence, the consideration of geometric resistance and flow distribution for the estimation of Windkessel model parameters is quite important, especially in a patient-specific geometry. The work done by S. Pant et al [18] considered the geometric resistance when estimating the Windkessel model parameters, but the resistances were approximated from an analytical formula ($\frac{8\mu L}{\pi R^4}$, where $\mu$ is the dynamic viscosity of fluid, $L$ is the length of the section, and $R$ is the radius of the vessel) by assuming the flow is a steady Poiseuille flow in a straight tube. The resistance obtained from this method may not reflect the real resistance under a patient-specific geometry and flow condition. Here, the geometric resistances calculated from the analytical formula for both geometries used in this study are presented in Table 3. In this calculation, we divided the branches into many smaller segments to ensure each segment has similar diameters and normal directions at two ends. The geometric resistances of each segment were calculated individually and added together accordingly to form the total resistance in each branch. Compared to Table 1, it is obvious that none of them can reflect the real geometric resistance. In general, the resistances are largely underestimated by this calculation. One obvious reason is that the Poiseuille flow formula does not consider the head losses due to bends and junctions in the aorta. Meanwhile, the resistance also depends on the flow pattern. For example, the $R_g$ of Case 3 (uniform flow distribution) is quite different from the ones of Case 1, as shown in Table 1. Inaccurate geometric resistance can affect the subsequent estimation of the WK3 parameters, causing errors in the final periodic CFD simulation.

Table 3. Geometric resistances calculated from the analytical equation by assuming the flow is Poiseuille flow. The two values for the DAo are geometric resistances of physiological geometry (left) and stenosed geometry (right) respectively. Unit: mmHg*s/mL.

| Branch | RSA | RCCA | LCCA | LSA | DAo |
|---|---|---|---|---|---|
| $R_g$ | 0.110 | 0.009 | 0.015 | 0.055 | 1.12E-3\|2.22E-3 |

It is worth mentioning that the estimation of WK3 parameters is under a reference pressure of 70 mmHg instead of the absolute pressure. The product of $R_p$ and $C$ from the WK3 determines the transient effect duration (relaxation time) of the electric circuit analogy. If the product is large, the ODE in the pattern search algorithm needs more iterations to reach the steady period. This transient effect does not matter much in the global optimization algorithm because of the low computational cost in the pattern search algorithm. However, a periodic CFD simulation may take a couple of hours to finish only one cardiac cycle. Furthermore, the driving force of an incompressible flow is the pressure drop between the inlet and outlets not the absolute pressure. Hence, we can use the reference pressure in the estimation of the Windkessel model parameters and the CFD simulation, which will give the same pressure waveform but reach a periodic solution faster due to a smaller relaxation time. Fig. 7 shows two pressure waveforms calculated from the ODE solver using absolute and reference pressures. These two curves almost overlap with each other. The corresponding WK3 parameters are listed in Table 4.

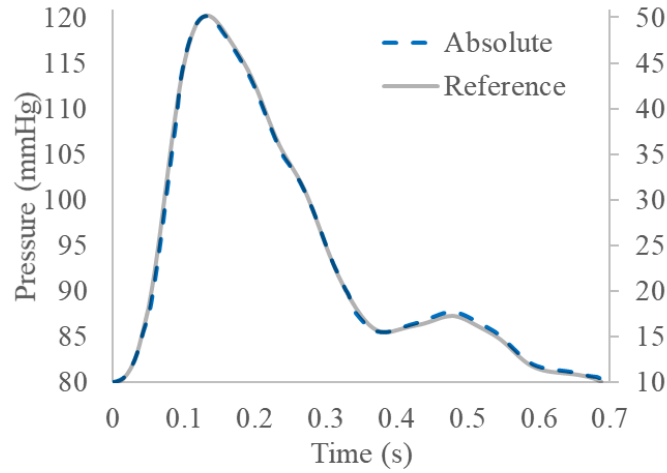

Figure 7. Pressure waveforms obtained from the ODE solver. Left y axis shows the pressure value of the absolute pressure case (dashed line) and right y axis denotes the pressure value of the reference case (solid line).

Table 4. WK3 parameters of the flow passing through the AAo. The target $P_{max}$ and $P_{min}$ are 120 and 80 mmHg for the absolute pressure case, 50 and 10 mmHg for the reference pressure case with $P_{ref} = 70$ mmHg. No geometric resistance is considered here.

| Case | $R_c$ (mmHg*s/mL) | $R_p$ (mmHg*s/mL) | C (mmHg/mL) | $R_pC$ (s) |
|---|---|---|---|---|
| Absolute | 0.187 | 1.518 | 2.314 | 3.512 |
| Reference | 0.182 | 0.244 | 2.079 | 0.508 |

In the algorithm, a steady CFD simulation with the mean flowrate during systole was used to find the geometric resistance. The reason to use a steady simulation is that the geometric resistance obtained from the steady CFD simulation with mean flowrate during systole is about the same as the one obtained from a periodic simulation, as shown in Table 5, where the geometric resistances in the periodic simulation were calculated using the mean pressure drop and mean flowrate over a cardiac cycle in Eq. 9. Note that the geometric resistances of the DAo in these two cases are quite different, but the values are small enough to neglect in comparison with the resistance from the WK3. Next, we examined the benefit of using the mean flowrate during systole as a characteristic flowrate compared to other flowrate choices. As shown in Table 6, all cases were obtained from CFD simulations with the normal geometry and evenly distributed flow. Case A ignored the geometric resistance, while Case B, C, and D used the mean flowrate during a whole cardiac cycle, mean flowrate during systole, and the peak systolic flowrate to find the geometric resistance, respectively. In Case A, the $L^2$ norm of error of flow distribution is the largest in all four cases, suggesting the importance of geometric resistances in regulating the flow distributions. In contrast, using the flowrate at peak systole (Case D) resulted in the largest geometric resistance. This case overestimated the geometric resistance, resulting in a larger error of Q%. Case C has the overall best result, which means the geometric resistance well reflects the flow resistance over the whole cardiac cycle. Hence, to avoid time-consuming periodic CFD simulations, a steady simulation with the mean flowrate during systole was used to find the geometric resistances.

Table 5. Geometric resistance of each downstream branch obtained from steady and periodic CFD simulations for Case 3. Unit: mmHg*s/mL.

| Case | DAo | RSA | RCCA | LCCA | LSA |
|---|---|---|---|---|---|
| Steady | 0.008 | 0.345 | 0.077 | 0.070 | 0.176 |
| Periodic | 4.51E-4 | 0.331 | 0.074 | 0.067 | 0.169 |

Table 6. Flow distribution and $L^2$ norm of error from CFD simulations of four different sets of WK3 parameters. Case A used a set of WK3 parameters without the consideration of geometric resistance, while Cases B, C, and D used the mean

flowrate during the whole cardiac cycle, the mean flowrate during systole, and the peak systolic flowrate to find the geometric resistance, respectively.

| Case | DAo | RSA | RCCA | LCCA | LSA | $L^2$ of Q% |
|------|--------|--------|--------|--------|--------|-------|
| A | 21.50% | 18.21% | 20.43% | 20.67% | 19.39% | 1.14% |
| B | 20.75% | 19.36% | 20.12% | 20.30% | 19.70% | 0.48% |
| C | 20.38% | 19.98% | 19.97% | 20.13% | 19.79% | 0.20% |
| D | 19.30% | 21.77% | 19.50% | 19.61% | 20.08% | 0.90% |

In the result section, Case 2 and 3 have slightly larger errors than Case 1 in terms of pressure. The resulting pressure waveforms at the AAo of all three cases are presented in Fig. 8a. The mean pressures of Case 1, 2, and 3 are 93.19, 93.41, and 93.68 mmHg respectively, which are very close to the target value of 93.33 mmHg (calculated by Eq. 12). However, the peak values of Case 2 and 3 overshoot slightly and the trough values undershoot, causing larger error in these two cases. There are two main sources relevant to this error. One source of error is that the flowrate waveforms at each outlet from CFD simulations are different from the estimated flowrate waveforms used in the ODE during the global optimization process. The flowrate waveforms of the DAo from both CFD and ODE input are shown in Fig. 8b, c, and d. As expected, the flowrate waveform has the largest discrepancy in Case 3. The waveforms of other branches follow similar behavior as the one of the DAo, so they are not presented in the figure. The flowrate waveforms used as the input of the ODE perfectly follow the shape of the inlet (AAo) flowrate waveform. However, in CFD simulations, the flowrate waveform was a result of the interaction between the pressure and the detailed flow pattern in each branch. Therefore, the shapes of flowrate waveforms at each outlet may deviate from the inlet one at the AAo. This issue might be solved by involving the iteration of periodic CFD simulations similar to Alimohammadi et al. [19] (We found that the error decreases from 4.13 to 3.41 mmHg after one iteration) or by adding another proper module in the circuit analogy to better mimic the flow phenomenon. As many clinical applications are time-sensitive, we need to make compromises between speed and accuracy. Therefore, the first time-consuming solution is not necessary for most applications since even for the extreme flow condition (Case 3), the error of the pressure is still acceptable. The second potential solution needs to be further studied and could be our future direction. The other source of error is the use of constant geometric resistance. The geometric resistance should vary with time since the flow pattern changes in a periodic manner. For example, the overall geometric resistance reaches the highest value during systole and decreases to a minimum during diastole. However, in this study, the geometric resistance is assumed to be constant. This assumption underestimates the resistance close to peak systole and overestimates that during diastole, finally inducing errors in pressure. Although the pressure waveform slightly diverges from the target value, the mean pressure is in excellent accordance with the target one, which indicates the geometric resistance obtained in this method can reflect the mean flow impedance during the whole cardiac cycle.

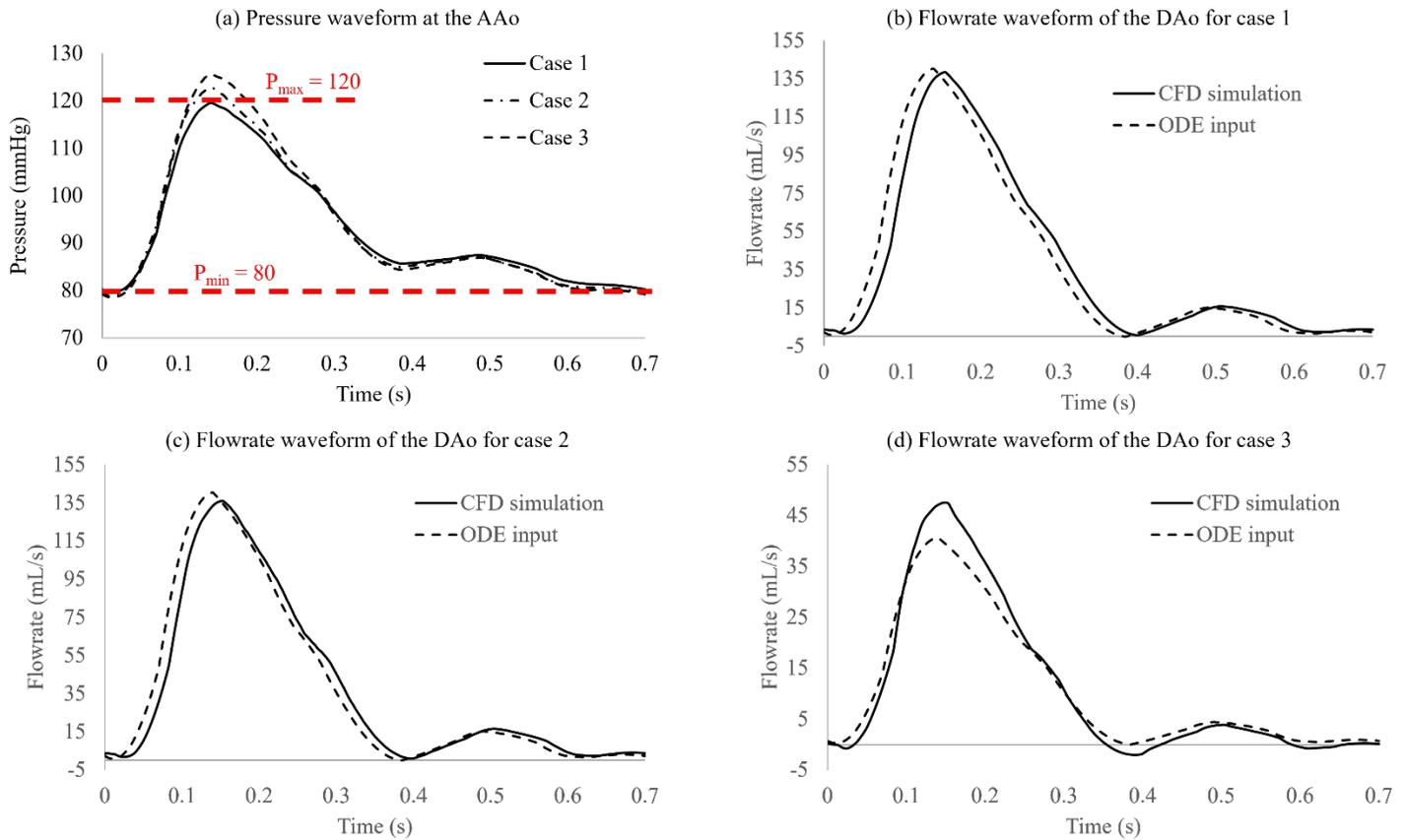

Figure 8. (a) shows the pressure waveform at the AAo of all three cases from CFD simulations with the denotation of the target pressure extremum. (b), (c), and (d) show the flowrate waveforms from the CFD simulation (solid curve) and the input of ODE (dotted curve) of Case 1, 2, and 3, respectively.

In this paper, several cases were examined, but only a single patient-specific geometry was used. Further simulations on different patient-specific geometries will allow for more evidence of the robustness of the algorithm. The flow distributions of downstream aortic branches were specified based on the data taken from the literature for healthy patients. More experiments should be conducted if other available patient-specific geometries and *in vivo* flowrate data are available. Besides the bridge to connect the resistance in CFD simulation and lumped parameter models presented in this paper, the relation of other modules between CFD simulations and lumped parameter models can also be a potential future direction. For example, adding compliances to mimic the behavior of flow in a moving wall CFD simulation can be a good direction and a good example to obtain the compliance (capacitor in an electric circuit) can be found in the literature from Bonfanti et al [21]. Furthermore, the validation with *in-vitro* or *in-vivo* data could be the future work of this study to provide more evidence on the robustness of this algorithm if relevant clinical data become available.

## 5. Conclusion

This paper proposed an algorithm to estimate the WK3 parameters for patient-specific multi-scale CFD simulations of aortic flow. With the known aorta geometry and inlet flowrate waveform, the proposed algorithm can fast estimate an optimized set of WK3 parameters for each outlet when coupled with CFD simulations, achieving the target extremum pressure at the AAo and target flow distributions of each outlet. The procedure involves two steps. In the first step, we conduct a steady CFD simulation to find an accurate set of geometric resistances for each branch. In the second step, a global optimization algorithm needs to be performed with the consideration of the geometric resistances in the circuit analogy. One advantage of this algorithm is the estimation procedure is remarkably fast. The global optimization algorithm takes only a couple of minutes to finish since only the ODEs need to be solved and the whole algorithm avoids the iteration of periodic CFD simulations. Another advantage is that our algorithm points out a potential way to get the mean geometric resistance accurately in a complex vascular system, such as the systems with cerebral aneurysms or aortic dissections. The algorithm is validated in a series of numerical experiments, including abnormal flow distributions, and artificially created stenosed geometry. The input variables of this study are easy to obtain from non-invasive clinical techniques. For example, the flowrate waveform may be obtained from echocardiography or phase-contrast MRI. The extremum aortic pressure can be acquired from a household sphygmomanometer. The patient-specific hemodynamics obtained from the multi-scale CFD model may better assist the clinicians with the prediction and diagnosis of cardiovascular diseases.

**Acknowledgement**

This work was supported in part by the American Heart Association 19CDA34660003 grant.